\documentclass[11pt]{article}
\usepackage{array}
\usepackage{graphicx}
\usepackage{cite}
\usepackage{textpos}

\title{Multiversality\footnote{Solicited review for Classical and Quantum Gravity}}

\author{Frank Wilczek\\
\small\it Center for Theoretical Physics, MIT, Cambridge MA 02139 USA}

\begin{document}

\maketitle

\begin{textblock*}{5cm}(11cm,-8.2cm)
  \fbox{\footnotesize MIT-CTP-4484}
\end{textblock*}

\begin{abstract}
Valid ideas that physical reality is vastly larger than human perception of it, and that the perceived part may not be representative of the whole, exist on many levels and have a long history.   After a brief general inventory of those ideas and their implications, I consider the cosmological ``multiverse'' much discussed in recent scientific literature.   I review its theoretical and (broadly) empirical motivations, and its disruptive implications for the traditional program of fundamental physics.  I discuss the inflationary axion cosmology, which provides an example where firmly rooted, plausible ideas from microphysics lead to a well-characterized ``mini-multiverse'' scenario, with testable phenomenological consequences.
\end{abstract}

\medskip

\section{Perspectives}

It is not difficult to design thought-experiments which demonstrate that scientists could easily form an inadequate idea of the extent of physical reality.   In his classic {\it Flatland} \cite{flatland}, Edwin Abbott imagined intelligent planar creatures, who infer a third dimension of physical space only after a baffling visit by a shape-shifting intruder.   The physicist and science fiction writer Robert Forward, in {\it Dragon's Egg} \cite{dragonsEgg}, imagined intelligent life evolving on a neutron star, effectively tied to its complex nuclear crust, and thereby provided a more-or-less physically plausible setting for the flatland scenario.  When human space-explorers visit their neutron star, these creatures must revise their cosmology.   (Note that neutron star life, based on nuclear rather than atomic processes, could be very quick-witted; Forward's story exploits that fact cleverly.)   Intelligent creatures evolved to live deep within the atmosphere of a gas giant planet could be deluded, for eons, into thinking that the Universe is an approximately homogeneous expanse of gas, filling a three-dimensional space, but featuring anisotropic  laws of motion (which we would ascribe to the planet's gravitational field).   

Are we human scientists comparably blinkered?

\subsection{Defining the Question}

Is there a multiverse, over and above the universe?  That meaning of that question, which is central to this review, is unfortunately confused in the ordinary usage of English language.   In my computer's dictionary  \cite{computerDictionary} ``universe'' is defined to mean ``all existing matter and space considered as a whole; the cosmos''.   Somewhat to my surprise, I found that ``multiverse'' is also defined in that dictionary.  It is defined to mean ``an infinite realm of being or potential being of which the universe is regarded as a part or instance''.   The authors of this dictionary have done an excellent job, both of keeping up-to-date and of drawing a subtle, nuanced distinction between those two concepts.   Nevertheless it is clear that tension between the two concepts remains.  In particular, if the ``universe'' contains everything that exists, what can be outside it?  If the answer is ``Things that don't exist'', then ``multiverse'' becomes an idea in the domain of psychology, not physics.     

For present purposes it is therefore useful to make some more precise -- though not, as will appear, entirely rigid -- distinctions.   By universe, I will mean the domain of physical phenomena which either are, or can reasonably be expected to be, accessible to observation by human beings in the foreseeable future.   By multiverse, I will mean a larger physical structure, of which the universe forms part.   Why should one entertain the possibility of a multiverse, if (by definition) it is unobservable?  As we shall discuss, there are several reasons.  Most importantly, it might be that the laws we use successfully to describe the observable universe are most naturally formulated in a larger framework, that includes unobservable parts.   Now purely as a logical matter it is always possible to append the dictum ``Nothing that is not observed exists.'' to any scientific doctrine, without changing its empirical content.  Indeed that dictum, which rules the multiverse out of court, is the essence of positivism, a recognized school of philosophy.  On the other hand, the dictum itself has no empirical content.   It is more in the nature of a moral exhortation, or methodological principle, not unrelated to Occam's razor and to Newton's {\it hypotheses non fingo}, whose application, though usually appropriate, might be outweighed by other scientific considerations.  

A second distinction, between related but more sophisticated notions, is also important.  The word ``universe'', as so far defined, carries no implication beyond observability.  But in modern scientific and cosmological discourse, it is often taken to mean more.   Specifically, it is tacitly assumed that the same fundamental laws apply at all places and at all times.  To highlight and separate that assumption, I will refer to it as ``universality'', and use ``Universe'' with an uppercase ``U'' to mean a universe that obeys universality.  (The ``Copernican Principle'' or ``Cosmic Mediocrity'' is a related, but not identical, idea.  It states, basically, that Earth does not occupy a privileged place in the universe.  Universality asserts more, namely that there are no privileged places or times.)  This suggests the coinage ``multiversality'', to mean that different laws apply at different places and/or times, and ``Multiverse'' to be a physical model embodying multiversality.   With this, it becomes possible to frame the sharp scientific question that will concern us here: 
\begin{quote}
Are there aspects of observable reality, i.e. the universe, that can be explained by multiversality, but not otherwise?
\end{quote}

The consensus thereby challenged, namely that physical reality is a Universe, exhibiting behavior that is uniform over all space and time -- i.e., that the universe obeys universality -- is a relatively new one, reached only after many surprising discoveries and much debate, as we shall review briefly below.  Even in its most modern, sophisticated embodiment, big bang cosmology, the idea of a Universe is subject to important qualifications.  The ``standard issue'' big bang Universe is spatially homogeneous only when averaged over exceedingly large scales, and then only approximately; also, it is far from homogeneous in time.   But those qualifications are well understood and easily taken into account.   The idea that there are unique physical laws that apply always and everywhere, which might be called universality, is the foundation of cosmology's standard model.  It has proved so powerful and successful that putting it into question might seem iconoclastic, frivolous, or both.  

Universality is also closely tied up with ideas of uniqueness and determinism, captured in such statements of faith as these by Spinoza \cite{spinoza}:
\begin{quote}
In nature there is nothing contingent, but all things are determined from the necessity of the divine nature to exist and act in a certain manner ...
\end{quote}
and Einstein \cite{einstein}:
\begin{quote}
I would like to state a theorem which at present can not be based upon anything more than upon a faith in the simplicity, i.e. intelligibility, of nature: there are no {\it arbitrary\/} constants ... nature is so constituted that it is possible logically to lay down such strongly determined laws that within these laws only rationally determined constants occur (not constants, therefore, whose numerical value could be changed without destroying the theory).
\end{quote}

We shall see, however, that there are now important reasons to question this consensus, and to favor some form of multiversality.

\subsection{Historical universes}

History affords valuable perspective on our questions.   

Pre-scientific ideas about the world, in general, implicitly adopt an attitude that the remarkable mathematician and philosopher Frank Ramsey, from a more sophisticated standpoint, made explicit: 
\begin{quote} 
My picture of the world is drawn in perspective and not like a model to scale. The foreground is occupied by human beings and the stars are all as small as three-penny bits. I don't really believe in astronomy, except as a complicated description of part of the course of human and possibly animal sensation. 
\end{quote} 
The viability of this point of view, we can now appreciate, is tied up with the locality of physical laws.  With a few notable exceptions the behavior of our immediate environment, in ways that affect everyday life, depends very little upon astronomical or cosmological phenomena.  The main exceptions are the day-night cycle, the seasons, and the tides; and these obey regularities that can be codified mathematically without reference to a physical theory of astronomy, as the Babylonians and Aztecs did.   The conjectured influence of planets on everyday life, codified in astrology, though empirically false, stimulated careful study of their motion, but again did not require, or fit with, the application of causal physical laws.   

A turn toward modern thinking occurred with the application of geometrical reasoning to astronomy in the work of Aristarchus, Eratosthenes and their followers, who calculated such things as the sizes of the Sun, Moon, and Earth, and their mutual distances.   These efforts presupposed the universal validity of geometry, and they established the smallness of Earth and human dimensions, relative to the size of the universe.   Later Greek and Hellenistic astronomy, however, did not adopt the idea of a Universe in our sense.   The Sun, Moon, and planets were supposed to be carried on Earth-centered spheres (later expanded to include spheres within spheres, or epicycles), as were the outermost ``fixed stars''.    Both the substance of these bodies -- a special fifth element or ``quintessence'' different from the terrestrial four -- and their laws of motion were, literally, unearthly.   This system, endorsed by philosophers and made quantitative (as a description of observed celestial phenomena, chiefly the motion of planets) by Hipparchus and Ptolemy, constituted a closed universe very different from (and much more serious than) Flatland.   It dominated educated opinion on astronomy and cosmology for several centuries, and even got connected to theological dogma.   

A dramatic series of discoveries overthrew that geocentric universe.  Copernicus proposed a simpler heliocentric system for planetary motion. Galileo's telescopic observations revealed unmistakable ``material'' and Earth-like features of the Moon ({\it e.g}. mountains and valleys), displayed a Jupiter-centric system of satellites manifestly analogous to Copernicus' model of the Solar System, and, by failure to resolve any structure, showed that stars must be very far away indeed, much more distant than the planets.   This, together with their brightness, gave credence to the idea, dating back to Aristarchus and perhaps even to the Pythagoreans, that the stars are Sun-like, and could support their own Earth-like worlds.   These developments, together with Kepler's accurate laws of planetary motion, came together in Newton's great synthesis.   Newton's classical mechanics lives most naturally in a Universe, since its laws are space- and time-translation invariant, and it draws no distinction in principle between terrestrial and celestial matter.   

Later, solar and stellar spectroscopy provided powerful evidence for universality, as common principles were found to govern all spectra, including those in terrestrial laboratories, the Sun, and distant stars \cite{spectroscopy}.   As one by-product, the hypothesis that the Sun is a typical star was vindicated.  

While these developments effectively established universality for physical law, they left open the question of whether matter is uniformly distributed, or whether we are embedded in an ``island universe'', our Galaxy, surrounded by void.   A key issue, not settled until the 1920s, was whether nebular objects, such as specifically the Andromeda Nebula, are local phenomena within our Galaxy, or very distant objects, constituting galaxies in their own right.   That this is not a trivial issue, is highlighted by the fact that many nebular objects, such as the Crab Nebula, the Orion Nebula, and others really are local.   It echoes earlier debates about the nature of stars.   Refinements in observational technique, including the use of Cepheid variables as standard candles and the availability of more powerful telescopes, resolved the issue in favor of uniform distribution, and still vaster cosmic dimensions.   

The overarching theme of this history is the emergence of a Universe, far vaster than human scales, but governed by universal laws, and roughly uniform in texture.   A relevant subtheme, is that in the past scientists have repeatedly reached ``intellectual closure'' on inadequate pictures of the universe, and underestimated its scale.  

\subsection{The Modern Universe}

Subsequent developments in cosmology, continuing to this day, have both refined the idea of a Universe, and provided impressive empirical support for it, from several lines of evidence.   I will be telegraphic in describing these developments, since they are the standard fare of cosmology textbooks and reviews \cite{cosmologyTexts}.  

The main {\it refinement}, already alluded to, is the realization that the Universe has evolved from a much denser, hotter, and more uniform state early in its history (big bang cosmology).  It is fair to call this a refinement, rather than a contradiction, of universality, because we use, with success, the same fundamental laws of physics to describe those early times.  

Major empirical successes of universality include:
\begin{itemize}
\item {\it uniformity -- surveys}: Deep surveys of galaxies show that their distribution approaches uniformity on scales above 150 Mpc or so.
\item {\it uniformity -- microwave background}: The existence of the microwave background, that accurately follows a blackbody spectrum, follows from the extrapolation of known laws of physics to the early universe.  Its accurate uniformity verifies the Universe hypothesis.
\item {\it nucleosynthesis -- origin of the elements}: The relative abundances of H$^1$, H$^2$, He$^3$, He$^4$ and Li$^7$ can be calculated by extrapolation of known laws of physics to big bang conditions.   The relative abundances of other isotopes can be calculated, with considerable quantitative precision, from the application of known laws of physics to stellar nucleosynthesis and ejection processes. 
\item {\it evolution of stellar populations}: The evolution of stars can be calculated from known laws of physics, and the ages indicated can be compared to other independent indicators, notably the measured expansion rate of the Universe (Hubble parameter).   The determinations are consistent.
\item {\it evolution of structure}: When viewing distant objects one is looking back in time, due to the finite speed of light.  There is a rich, successful theory of how structures such as galaxy clusters, galaxies themselves, and stars evolved from an initially very uniform mass distribution (as indicated by the microwave background) through gravitational instabilities.
\end{itemize}

\subsection{Cohabiting Worlds}

A common habit of thought, implicit in much of the preceding discussion, is the idea that space is simple receptacle in which bodies move around, with no two bodies present at the same point.  This space-concept is deeply embedded in human psychology, as it underlies our usual interpretation of the visual world.  We construct, from primitive light-perceptions, a world of objects moving in space.   It is also the dominant space-concept in classical mechanics, where Newton envisioned atomic ``hard, massy, {\it impenetrable\/} spheres'' (emphasis added).  

In modern quantum physics generally, and in the standard model of fundamental physics in particular, physical space appears as a far more flexible framework.  Many kinds of particles can be present at the same point in space at the same time.   Indeed, the primary ingredients of the standard model are not particles at all, but an abundance of quantum fields, each a complex object in itself, and all omnipresent.

The framework of quantum theory allows, and seems to demand, that we envisage ourselves within a multiverse of a qualitatively new kind.  The traditional ``cosmological'' Multiverse considers that there might be physical realms inaccessible to us due to their separation in space-time.  The quantum Multiverse arises from entities that occupy the same space-time, but are distant in Hilbert space -- or in the jargon, decoherent.   

The basic idea of decoherence can be illustrated simply in a toy example, as follows.  Consider the wave function $\psi (x_1, x_2, ... , x_N)$ of a system of particles, at a fixed time.  Let us suppose that it decomposes into two pieces 
\begin{eqnarray}
\psi (x_1, x_2, ... , x_N) ~&=&~ \phi_1 (x_1, ... , x_k) f(x_{k+1}, ... , x_N) \nonumber \\
&+&~  \phi_2 (x_1, ... , x_k) g(x_{k+1}, ... , x_N)
\end{eqnarray}
with 
\begin{eqnarray}
\int \, \prod\limits_{j=k+1}^N \, dx_j \, f^*(x_{k+1}, ... , x_N) g(x_{k+1}, ... , x_N) ~=~ 0 \nonumber \\
\int \, \prod\limits_{j=k+1}^N \, dx_j \, f^*(x_{k+1}, ... , x_N) f(x_{k+1}, ... , x_N) ~=~ 1 \nonumber \\
\int \, \prod\limits_{j=k+1}^N \, dx_j \, g^*(x_{k+1}, ... , x_N) g(x_{k+1}, ... , x_N) ~=~ 1 
\end{eqnarray}
Then the expectation value of an observable ${\cal O}(x_l)$ that depends only on the $x_l$ with $1 \leq l \leq k$ will take the form
\begin{eqnarray}
\langle \psi | {\cal O} (x_l) | \psi \rangle ~&=&~ \int \, \prod\limits_{j=1}^N \, dx_j \, \psi^*(x_1, ... , x_N) {\cal O} (x_l) \psi (x_1, ... , x_N) \nonumber \\
~&=&~ \int \, \prod\limits_{j=1}^k \, dx_j \, \phi_1^*(x_1, ... , x_k) {\cal O} (x_l) \phi_1 (x_1, ... , x_k) \nonumber \\ 
&+&~ \, \int \, \prod\limits_{j=1}^k \, dx_j \, \phi_2^*(x_1, ... , x_k) {\cal O} (x_l) \phi_2 (x_1, ... , x_k)
\end{eqnarray}
Thus there is no communication between the branches of the wave function based on $\phi_1$ and $\phi_2$.  In this precise sense those two branches describe mutually inaccessible (decoherent) worlds, both made of the same materials, and both occupying the same space.   

It can be shown that all but the simplest (or most carefully crafted) quantum-mechanical systems rapidly evolve, generically, into wave functions with many decoherent branches.  In particular ``measurement processes'', wherein macroscopic entities settle stably into one outcome or another from among a menu of distinct possibilities, generate decoherent branches corresponding to the different outcomes.   According to the many-worlds interpretation of quantum mechanics (as I understand it), this provides a possible way to reconcile the probabilistic nature of quantum prediction with the mathematically unique (= deterministic) evolution of states according to the Schr\"odinger equation.   The probabilities assigned to different outcomes, in this interpretation, describe the probabilities for finding oneself on the branch of the wave function, or ``world'', where the given outcome has occurred.   The wave function as a whole evolves deterministically, but it describes a multiverse, only one part of which remains accessible.  

This is an important example of multiversality, which should be (but is not) uncontroversial.  It provides a positive answer to our question 
\begin{quote}
Are there aspects of observable reality, i.e. the universe, that can be explained by multiversality, but not otherwise?
\end{quote}
in the form 
\begin{quote}
Yes -- one is the apparent indeterminism of quantum mechanics, despite its deterministic equations.
\end{quote}

\section{The Multiverse in Modern Cosmology}

\subsection{Arguments and Evidence}

Another potential source of multiversality is the more straightforward possibility that space is much larger, in the ordinary sense of geometry, than the observable universe.  A weak form of this is already empowered by conventional big bang cosmology, which gives rise to an ever-expanding horizon -- new regions of space open to view as the universe ages, since light can have traveled further.  Less conventional is the possibility that distant regions exhibit radically different content or behaviors from the observed (universal) universe.   I think it is fair to say that that second view, while it had serious advocates \cite{tiplerBarrow}, was sparsely represented within the physics literature until the early twenty-first century.   Yet by now it is widely accepted as conventional wisdom.  What happened?  

This is not the place to consider the sociology of the conversion process, although that is both interesting and explanatory.   Several intellectual developments contributed to the change, and might justify it intellectually:

\begin{description}
\item[1. The standardization of models:] 

With the extraordinary success of the standard model of fundamental physics, brought to a new level of precision at LEP through the 1990s; and with the emergence of a standard model of cosmology, confirmed by precision measurements of microwave background anisotropies, it became clear that an excellent working description of the world as we find it is in place.  In particular, the foundational laws of physics that are relevant to chemistry and biology seem pretty clearly to be in place.  

The standard models are founded upon broad principles of symmetry and dynamics, assuming the values of a handful of numerical parameters as inputs.   Given this framework, we can consider in quite an orderly way the effect of a broad class of plausible changes in the structure of the world: namely, change the numerical values of those parameters!   When we try this we find, in several different cases, that the emergence of complex structures capable of supporting intelligent observation appears quite fragile.  

On the other hand, valiant attempts to derive the values of the relevant parameters, using symmetry principles and dynamics, have not enjoyed much success.   

Thus life appears to depend upon delicate coincidences that we haven't been able to explain.  The broad outlines of that situation have been apparent for many decades.  When less was known, however, it seemed reasonable to hope that better understanding of symmetry and dynamics would clear things up.  Now that hope seems much less reasonable.  The happy coincidences between life's requirements and nature's choices of parameter-values might be just a series of flukes, but one could be forgiven for beginning to suspect that something deeper is at work.   

That suspicion is the first deep root of anthropic reasoning. 

\item[2. The phase transition paradigm:]

The standard model of fundamental physics incorporates, as one of its foundational principles, the idea that ``empty space'' or ``vacuum'' can exist in different phases, typically associated with different amounts of symmetry.    Moreover, the laws of the standard model itself suggest that phase transitions will occur, as functions of temperature.   Extensions of the standard model to build in higher symmetry (gauge unification or especially supersymmetry) can support effective vacua with radically different properties, separated by great distance or by domain walls.   That would be a form of failure of universality, in our sense, whose existence is suggested by the standard model.

\item[3. The exaltation of inflation:]

As previous emphasized, the most profound result of observational cosmology, as emphasized previously, has been to establish the Universe, in which the same laws apply everywhere and everywhen, and moreover matter is, on average, of the same kind and uniformly distributed throughout.   It would seem only reasonable, then, to think that the observed laws are unique, allowing no meaningful alternative, and to seek a unique explanation for each and every aspect of them.   Within that framework, explanations of basic laws, properties of matter, or cosmography that invoke selection effects are moot.  If there is no variation, then there cannot be selection. 

Inflationary cosmology challenges that inference.  It proposes a different explanation of universality.  According to inflationary cosmology, the observed universe originated from a small patch, and had its inhomogeneities ironed out dynamically.   In most theoretical embodiments of inflationary cosmology, the currently observed universe appears as a small part of a much larger multiverse.   In this framework to hold throughout the universe need not hold through all space.  They can be accidents of our local geography, so to speak.   If that is so, then it is valid -- indeed, necessary --  to consider selection effects.  It may be that some of the ``fundamental constants'', in particular, cannot be determined by theoretical reasoning, even in principle, because they really are different elsewhere.   

The success of inflationary cosmology \cite{inflationReview} is the second deep root of anthropic reasoning.

\item[4. The unbearable lightness of space-time:]

Modern theories of fundamental physics posit an enormous amount of structure within what we perceive as empty space: quantum fluctuations, quark-antiquark condensates, Higgs fields, and more.   At least within the framework of general relativity, gravity responds to every sort of energy-momentum, and simple dimensional estimates of the contributions from these different sources suggest values of the vacuum energy, or cosmological term, many orders of magnitude larger than what is observed.   Depending on your assumptions, the discrepancy might involve a factor of $10^{60}, 10^{120}$, or $\infty$.  (The first of these estimates derives its energy scale from the electroweak scale, perhaps associated with low-energy supersymmetry; the second from breaking of unified gauge symmetry, the third from the divergent zero-point energy of generic quantum field theories.)

Attempts to derive an unexpectedly small value for this parameter, the vacuum energy, have not met with success.   Indeed most of those attempts aimed to derive the value zero, which now appears to be the wrong answer.  

In 1987 Weinberg proposed to cut the Gordian knot by applying anthropic reasoning to Einstein's cosmological term, or in modern usage the dark energy density $\rho_{\rm DE}$ \cite{weinberg}.  On this basis he predicted that the dark energy density, rather than being zero, would be as large as it could be, while remaining consistent with the emergence of observers.  When observations coalesced around a substantial value of $\rho_{\rm DE}$ \cite{darkEnergy}, this proposal gained enormous credibility.

The numerical accuracy of the anthropic prediction is not overwhelmingly impressive (the computed probability to observe a cosmological term as small as we do is roughly 10\%), though that might be laid to the vagaries of sampling a statistical distribution just once and uncertainties in applying the selection counterfactual.   More fundamentally, the calculation is based on the hypothesis that one should consider variations in the vacuum energy alone, keeping all other parameters fixed, which is a questionable assumption.   Indeed the parameter that most naturally appears in estimating selection effects is not directly the dark energy density $\rho_{\rm DE}$ but the combination 
\begin{equation}
p ~\equiv~ \frac{\rho_{\rm DE}}{\xi^4 Q^3}
\end{equation}
where $\xi$ is the late-time mass density of matter per photon and $Q$ is the amplitude of density fluctuations.   One needs $p < 1$ to form dark matter halos, and to form structures that plausibly support observers.  Now $Q$ is observed to be rather small ($\sim 10^{-4}$), while the measure in microphysical model-space plausibly opens up sharply toward higher $\rho_{\rm DE}$, this criterion begs the question of why {\it both\/} of those parameters are not larger, for an optimal $p$.  
Nevertheless, the apparent observation of vacuum energy that is ridiculously small from a microphysical perspective, but importantly large from a cosmological perspective, and the current lack of viable alternatives, certainly encourages one to take the explanation based on selection seriously.    It gives us a second answer to our question:
\begin{quote}
Are there aspects of observable reality, i.e. the universe, that can be explained by multiversality, but not otherwise?
\end{quote}
in the form 
\begin{quote}
Yes -- the outrageously small, but non-zero, value of the dark energy density.   
\end{quote}

\item[5. The superabundance of string theory:]

After a brief, heady period around 1984-5, during which it seemed that simple general requirements (e.g., $N=1$ supersymmetry and three light fermion generations) might pick out a unique Calabi-Yau compactification as the description of observed reality, serious phenomenological application of string theory has been forestalled by the appearance of a plethora of candidate solutions.    The solutions all exhibited unrealistic features (e.g. unbroken supersymmetry, extraneous massless moduli fields), and it was anticipated that when those problems were fixed some degree of uniqueness might be restored.    It was also hoped that string theory would provide a dynamical understanding for why the cosmological term is zero \cite{GSW}.  

Recent constructions have provided a plethora of approximate solutions with broken supersymmetry and few or no moduli fields.   They are not stable, but it is plausible that some of them are metastable with very long lifetimes indeed.   As yet none (among $\geq 10^{\rm hundreds}$) appears to be entirely realistic, but there's still plenty of scope for investigation in that direction, and even for additional constructions.    

In these new constructions the cosmological term can take a wide range of values, positive or negative.   So if cosmology provides a Multiverse in which a significant sample of these metastable solutions are realized, then the stage might be set for selection effects to explain (roughly) the value we actually observe by anthropic reasoning, as just sketched.

\end{description}

Many other discussions of anthropic reasoning, to supply explanations of other parameters, have appeared in the literature.  Life in anything close to the form we know it requires both that there should be a complex spectrum of stable nuclei, and that the nuclei can get synthesized in stars.  As emphasized by Hogan \cite{hogan}  and many others,  those requirements imply constraints, some quite stringent, relating the QCD parameters $\Lambda_{\rm QCD}, m_u, m_d$ and $m_e$ and $\alpha$.   On the other hand these parameters appear on very different footings within the standard model and in existing concrete ideas about extending the standard model.   The required conspiracies {\it among\/} the masses $m_u, m_d, m_e$ are all 
the more perplexing because each of the masses is far smaller than the ``natural'' value, ~250 GeV, set by the Higgs condensate.  An objective measure of how unnatural this is, is that pure-number Yukawa couplings of order $10^{-6}$ underlie these masses.   

More recent is the realization that the emergence of observer-friendly macrostructures, that is stable planetary systems, requires rather special relationships among the parameters of the cosmological standard model.   Here again, no conventional symmetry or dynamical mechanism has been proposed to explain those relationships; indeed, they connect parameters whose status within existing microscropic models is wildly different.  Considerations of this sort have a rich literature, beginning with \cite{carr}.

Less emphasized, but to me also highly significant, is the abundance of standard model parameters whose values are {\it not\/} connected to the emergence of observers in any obvious or even plausible way, and which have proved notoriously resistant to theoretical understanding. These include the masses and weak mixing angles  of the heavier quarks and leptons (encoded in the Cabibbo-Kobayashi-Maskawa, or CKM, matrix), and the masses and mixing angles of neutrinos.  It also includes most of the prospective parameters of models beyond the standard model, such as low-energy supersymmetry, because only a few specific properties of those models (e.g., the rate of baryogenesis) are relevant to late-time physics, let alone life.  

The point is that a cosmological multiverse that supports enough variation to allow selection to operate among a significant fraction of the parameters that are relevant to life, will also allow variation among parameters that are not relevant to life.  This gives us a third answer to our question:
\begin{quote}
Are there aspects of observable reality, i.e. the universe, that can be explained by multiversality, but not otherwise?
\end{quote}
in the form 
\begin{quote}
Yes -- the opaque and scattered values of many standard model parameters that are {\it not\/} subject to the discipline of selection.  
\end{quote} 

\subsection{Prospect}

It seems appropriate to close this Section with a lamentation and a warning.  
\begin{itemize}
\item {\it Lamentation}: I don't see any realistic prospect that anthropic or statistical selection arguments -- applied to a single sample! -- will ever lead to anything  comparable in intellectual depth or numerical precision to the greatest and most characteristic achievements of theoretical physics and astrophysics, such as (for example) the prediction of electron and muon anomalous magnetic moments, the calculation of the hadron spectrum, or the enabling of GPS, celestial navigation, and interpretation of pulsar timing.  In that sense, intrusion of selection arguments into foundational physics and cosmology really does, to me, represent a lowering of expectations. Moreover, because the standard models of fundamental physics and cosmology describe the world so well, a major part of what ideas going beyond those standard models could aspire to achieve, for improving our understanding of the world, would be to fix the values of their remaining free parameters.  If we compromise on that aspiration, there will be fewer accessible features of the physical world for fundamental theory to target.  One sees these trends, for example, in the almost total disconnect between the subject matter of hep-th and hep-ex. 
\item {\it Warning}:  There is a danger that selection effects will be invoked prematurely or inappropriately, and choke off the search for deeper, more consequential explanations of observed phenomena.  To put it crudely, theorists can be tempted to think along the lines ``If people as clever as us haven't explained it, that's because it can't be explained -- it's just an accident.''   I believe there are at least two important regularities among standard model parameters that {\it do\/} have deeper explanations, namely the unification of couplings and the smallness of the QCD $\theta$ parameter (for which, see below).  There may well be others.  
\end{itemize}

\bigskip

\section{Inflationary Axion Cosmology}

\subsection{Principles}

\subsubsection{Microphysical Principles}

The theory of the strong interaction (QCD) admits a parameter, $\theta$, that is observed to be unnaturally small: $|\theta | < 10^{-9}$.   That suspicious ``coincidence'' can be understood by promoting translation of $\theta$ to an asymptotic or classical quasi-symmetry, Peccei-Quinn (PQ) symmetry.    

The cosmological considerations that follow depend only on some relatively simple consequences of PQ symmetry, not on the rather subtle aspects of quantum field theory that motivate the symmetry and lead to those consequences.  Nevertheless, a few (optional) comments on those subtleties seem in order.

The $\theta$ term in QCD is a possible interaction of the type
\begin{equation}\label{thetaInteraction}
{\cal L} ~=~ \frac{\theta}{32\pi^2} \ {\rm Tr} \, \epsilon^{\alpha \beta \gamma \delta} \, G_{\alpha \beta} \, G_{\gamma \delta}  
\end{equation}
where $G_{\alpha \beta}$ is the field strength for the gluon field of QCD.  (Here the normalization of the gauge potential has been chosen so that the coupling constant does not appear in covariant derivatives, but only in the coefficient of the Yang-Mills kinetic term.)  This interaction is unusual in several respects. 
\begin{enumerate}
\item The parameter $\theta$ is a pure number.  The interaction of Eqn.\,(\ref{thetaInteraction}) is therefore, in the language of quantum field theory, ``strictly renormalizable'' or ``marginal''.  In particular, it is not sequestered from effects of interactions at very large mass scales, by inverse powers of the scale.   
\item Formally, ${\rm Tr} \, \epsilon^{\alpha \beta \gamma \delta} \, G_{\alpha \beta} \, G_{\gamma \delta}$ is a total divergence, and so the $\theta$ term does not contribute to the equations of motion.  However the quantity of which it is the divergence is not locally gauge invariant, and it can therefore contain singularities.   The integrated Lagrangian density, or action, is sensitive to the singular surface terms which arise upon integration by parts.  Since configurations in the Feynman functional integral, which defines the quantum theory, are weighted by their action, the quantum theory (as opposed to the classical theory) does depend on $\theta$.   
\item A refined analysis of the allowed singularities shows that $\theta$ is periodic, with period $2\pi$, in the sense that physical results only depend on the value of $\theta$ modulo $2\pi$. 
\item Under the discrete symmetries of parity $P$ and time reversal $T$ we have
\begin{equation}
\theta ~\stackrel{P, T}{\longrightarrow}~ - \theta
\end{equation}
In view of the periodicity of $\theta$, we see that $P$ and $T$ invariance requires $\theta \equiv 0 $ modulo $\pi$. 
\end{enumerate}

If $\theta \neq 0$ modulo $\pi$, we will have an interaction within QCD that violates $P$ and $T$, but no other symmetries.  Such an interaction would reveal itself physically in the existence of a non-zero electric dipole moment for the neutron.   There are powerful constraints on the magnitude of such a moment, which essentially translate into the bound $|\theta | < 10^{-9}$ mentioned earlier.  Since $\theta$ is periodic, this is to be understood as a bound on $\theta$ modulo $2\pi$.  Values of $|\theta |$ near $\pi$ modulo $2\pi$, though they respect $P$ and $T$, are excluded for other reasons.

Physicists have become accustomed to using various kinds of approximate symmetry, some quite subtle, in the description of nature.  Gauge symmetry in QCD, for example, is a statement about how the theory is formulated, rather than about physical processes directly: all physical states are gauge singlets.   PQ ``symmetry'' postulates that there is a transformation depending on a single real parameter $\lambda$ whose net effect is to modify the Lagrangian density of the world according to
\begin{equation}\label{lagrangianChange}
\Delta {\cal L} ~=~ \frac{\lambda}{32\pi^2} \ {\rm Tr} \, \epsilon^{\alpha \beta \gamma \delta} \, G_{\alpha \beta} \, G_{\gamma \delta}  
\end{equation} 
(It is possible to loosen this postulate somewhat, allowing additional terms of a similar kind but involving the $SU(2)\times U(1)$ gauge fields to occur on the right-hand side.  Such terms arise in many specific models incorporating PQ symmetry.  They affect some significant details of axion phenomenology, but not the main story-line that follows.)   This change has the effect of modifying the $\theta$-term of QCD, according to 
\begin{equation}\label{thetaChange}
\Delta \theta ~=~ \lambda
\end{equation}

Putting off for a moment why it is desirable, let us ask: Why is the existence of a relationship like Eqn.\,(\ref{lagrangianChange}) theoretically plausible?  After all, conventional symmetry statements are not quite of this form: They have a vanishing right-hand side!   Here we have not a vanishing right-hand side, but rather a specific, non-zero change in $\cal L$.  So the question is: What is special about this particular form of change?  In fact the right-hand side of Eqn.\,(\ref{lagrangianChange}) has several special features, which additional terms would spoil.  One is that it is a uniquely ``soft'' interaction: its effects diminish rapidly at high momentum transfers, or at high temperature.   Another, related property is that its effects vanish in the classical limit, and to all orders in perturbation theory.  Both of those features are consequences of the fact that it is a total divergence.   So if we are ready to postulate not only exact symmetries, but also classical or asymptotic symmetries, then PQ `symmetry' qualifies.

It is also possible the PQ symmetry could arise ``accidentally'', as an indirect consequence of other principles, rather than as an independent principle.   This possibility is accentuated by the nature of PQ symmetry, that it is a symmetry of the classical Lagrangian.   Indeed, the classical Lagrangians associated with conventional quantum field theories, being restricted to low-order polynomials in the fields, are far from being the most general functions of their variables, and may not be able to supply the non-singlet terms that would break a candidate PQ symmetry.  

At a technical level, the relationship of Eqn.\,(\ref{lagrangianChange}) defines an {\it anomalous\/} symmetry.  When a classical theory with PQ symmetry is quantized, the right-hand side arises from a very specific class of Feynman graphs, namely the triangle graphs whose vertices connect two color gluons to an insertion of the PQ symmetry current.  It is an important result of quantum field theory, that these, and no other, correction terms arise.   Models based on string theory can also naturally incorporate PQ symmetry, through a variety of mechanisms \cite{stringAxion}.

Now let us discuss why the postulate of Eqn.\,(\ref{lagrangianChange}) is desirable.  Let us suppose that there is a complex scalar field $\phi$, which may be fundamental or composite, which transforms as 
\begin{equation}\label{phiTransformation}
\phi^\prime ~=~ e^{i\lambda} \phi  
\end{equation}
and that acquires a vacuum expectation value of magnitude 
\begin{equation}
| \langle \phi \rangle | ~=~ F
\end{equation}
The possible values of the phase of $\langle \phi \rangle $ will, according to Eqn.\,(\ref{phiTransformation}), correspond to different values of $\lambda$, and thereby, through Eqn.\,(\ref{thetaChange}), to different values of $\theta$, leaving the remainder of the Lagrangian density unchanged.  But changes in $\theta$ have a very specific effect on the vacuum energy density, related to instantons in QCD, which can be calculated.   The values of $\theta$ where the discrete symmetries $P$ and $T$ are valid, i.e. $\theta \equiv 0 $ modulo $\pi$, are points of enhanced symmetry, and can be expected to be stationary points of the vacuum energy.  Detailed calculations bear out that expectation, and furthermore suggest that $\theta \equiv 0$ modulo $2\pi$ gives the minimum energy.    The upshot of all this is that the observed, effective value of the $\theta$ parameter in the physical ground state will be very small, for dynamical reasons.  (It is not quite zero, since $P$ and $T$ are, at the relevant low energies, slightly broken symmetries.)   That result is consistent with the otherwise apparently ``unnatural'' value of $\theta$ which is observed, and thus potentially explains that puzzling feature of the world.

The axion field $a$ is established at the Peccei-Quinn transition, when our complex order-parameter field $\phi$ acquires an expectation value $F$:
\begin{equation}
\langle \phi \rangle ~=~ F e^{i\theta} ~=~ F e^{ia/F}
\end{equation}
This form is chosen so that the kinetic energy term for the $a$ field, 
\begin{equation}
\partial^\mu \phi \, \partial_\mu \phi ~\stackrel{\tilde{}}{\rightarrow}~ \partial^\mu a \, \partial_\mu a
\end{equation}
is conventionally normalized.  At the level of the classical Lagrangian we have have a shift symmetry $a \rightarrow a + \lambda$, which forbids mass terms for $a$.  There is, however, energy associated with variation in the magnitude of $a$, arising from the right-hand side of Eqn.\,(\ref{lagrangianChange}).   Changes in $a$ effectively generate changes in the $\theta$ parameter, according to 
\begin{equation}\label{aThetaRelation}
\delta \theta ~=~ \delta \frac{a}{F}
\end{equation}
Since changes in $\theta$ by finite angles are associated with significant changes in the QCD vacuum, we might expect that the total energy density $\cal E$ in play is of order $\Lambda_{\rm QCD}^4$, where $\Lambda_{\rm QCD}$ is a typical QCD scale, say $\sim 100$ MeV.   Assuming for simplicity a minimal trigonometric form for the functional ${\cal E} (\theta)$, viz.
\begin{equation}
{\cal E} (\theta ) ~\approx~ \Lambda_{\rm QCD}^4 (1 - \cos \theta)
\end{equation}
and expanding around $\theta = 0$, using Eqn.\,(\ref{aThetaRelation}), we arrive at the effective mass$^2$ term
\begin{equation}\label{massEstimate}
m^2 ~=~ \frac{d^2 \cal E}{da^2} ~=~ \frac{\Lambda_{\rm QCD}^4}{F^2}
\end{equation}
This estimation can be done much more accurately, but for present purposes it is sufficient to note its major qualitative features, evident in Eqn.\,(\ref{massEstimate}):
\begin{itemize}
\item The mass is {\it inversely\/} proportional to the symmetry-breaking scale $F$.
\item Its order of magnitude is 
\begin{equation}
m_a ~\approx~ \frac{(100 \ {\rm MeV})^2}{F}
\end{equation}
Thus for $F = 10^{12}$ GeV we have $m_a ~\approx~ 10^{-5}$ eV, and for $F = 10^{16}$ GeV $m_a ~\approx~ 10^{-9}$ eV.  (We shall soon see that these are plausible $F$ values.) Thus $a$ is predicted to be an exceeding light particle. 
\end{itemize}

\subsubsection{Cosmological Principles}

At temperatures $T >> F$, the vacuum expectation value of the $\phi$ field will vanish.   On dimensional grounds, we should expect a phase transition, wherein $\phi$ acquires a vacuum expectation value, at $T \sim F$.  At the time of transition, which occurs (if at all) in the very early universe, the energy associated with varying $a_0$, or equivalently $\theta_0$, is negligible, both because there are much larger energy densities than $\Lambda_{\rm QCD}^4$ in play, and because high temperature suppresses the dependence of the energy density on $\theta$.  Thus differences from the minimum $a =0$, arising from the stochastic nature of the phase transition, can be imprinted.  The locally imprinted value persists almost unchanged until it becomes energetically significant, at $T \sim \Lambda_{\rm QCD}$.   At that point, the scalar field $a_0$ diminishes in magnitude, and materializes as a Bose-Einstein condensate of axions.  A standard analysis, which I will not reproduce here, shows that the consequent mass density today is roughly proportional to $F\sin^2 \theta_0$.  

If no inflation occurs after the Peccei-Quinn transition then the spatial correlation length in this mass density, which by causality was no larger than the horizon when the transition occurred, corresponds to a very small length in the present universe.  To calculate the axion contribution the mass density of present-day universe on cosmological scales, therefore, we should simply average over $\sin^2 \theta_0$.  One finds that $F \sim 10^{12}$ GeV corresponds to the observed dark matter density \cite{aCosmo1} \cite{aCosmo2} \cite{aCosmo3}. 

Since experimental constraints require \cite{axionConstraints} $F \geq 10^{10}$ GeV, axions are almost forced to be an important component of the astronomical dark matter, if they exist at all.  So it seems interesting to entertain the hypothesis that axions provide the bulk of the dark matter, and $F \cong 10^{12}$ GeV.  That has traditionally been regarded as the default axion cosmology.  A cosmic axion background with $ F \cong 10^{12}$ GeV might be detectable, in difficult experiments.  Searches are ongoing, based on the conversion $a \rightarrow \gamma \gamma (B)$ of axions into microwave photons in the presence of a magnetic field \cite{conventionalAxionExperiments}.    

If inflation occurs after the Peccei-Quinn transition, things are very different.  In that scenario a tiny volume, which was highly correlated at the transition, inflates to include the entire presently observed universe.  So we shouldn't average.  As a result, $F > 10^{12}$ GeV can be accommodated, by allowing ``atypically'' small $\sin^2 \theta_0$ \cite{aCosmo1}.  

But now we must ask, by what measure should we judge what is ``atypical''?  In the large-$F$ scenario, most of the multiverse is overwhelmingly axion-dominated, and inhospitable for the emergence of complex structure, let alone observers.  Thus it is logically justified, and methodologically appropriate, to consider selection effects \cite{linde}.

\subsection{Results}

In the large-$F$, inflationary axion cosmology, $\theta_0$ controls the dark matter density, but has little or no effect on anything else.   We therefore have a direct relationship between a random variable with a definite distribution and a single, easily interpreted physical parameter.   That puts us in an ideal position to work within a well-characterized multiverse, correct for selection effects, and estimate the relative probability that the value of that physical parameter -- i.e., the dark matter density -- is a probable one.   It is hard to imagine a clearer, cleaner case for applying anthropic reasoning. 

Although we have escaped several of the severe limitations and difficulties of anthropic reasoning mentioned previously, two major ones remain.  One is that we must somehow bridge the gap between a physically specified multiverse, whose properties are (statistically) well-defined in space and time, to an account of what the ``typical observer'' sees.  The other, of course, is that at the end of the day our sample seize is still one.   Even if -- as turns out to be the case -- we find that the dark matter density seen by a ``typical observer'' is a random variable with a reasonably peaked and reasonably narrow distribution, our prediction should be assessed in that light.

The astrophysical theory of structure formation, supplemented by standard cosmological initial conditions on fluctuations in dark matter and baryon densities to wit a roughly scale-invariant adiabatic spectrum with the ratio of densities 
\begin{equation}\label{darkMatterAbundance}
r ~\equiv~ \frac{\rho_{\rm DM}}{\rho_{\rm baryon}} ~\approx~ 6
\end{equation}
gives a successful account of the broad features of galaxy formation.  This includes, notably, predicting the existence of ``observer-friendly'' galaxies -- that is, galaxies which neither collapse to gigantic black holes, nor remain as diffuse (non-stellar) gas clouds.    As one contemplates varying the ratio in Eqn.\,(\ref{darkMatterAbundance}), the same theory of structure formation indicates that these and other difficulties arise, which seem to pose formidable difficulties for the emergence of intelligent observers.  

To quantify this observation, in \cite{tarw} Tegmark, Aguirre, Rees and I defined a set of cuts to specify the notion of ``observer-friendly'' structures, and within those observer-friendly structures took the number of baryons as a rough-and-ready measure of the number of observers likely to arise.  With these (admittedly crude) definitions in place, we could calculate the probability distribution for the ratio $r$, {\it per observer}.  The result of our analysis is encouraging.  Taken at face value, it suggests that in the large $F$ axion cosmology the typical observer sees a ratio of dark to baryonic matter close to what we observe in our neighborhood (that is, in the universe visible to us!).

\subsection{Prospect}

The reasoning leading to axions and the cosmology they suggest is unusually long and intricate, even by the standards of theoretical physics, but each step has survived extensive scrutiny.  As we have seen, the large-$F$ inflationary axion scenario provides a uniquely clear, clean showcase for anthropic reasoning, leading to a successful (though loose) prediction for $r$, the relative dark matter abundance.   

Discovery of large-$F$ axions would provide powerful evidence for a cosmological multiverse (as well as solving the fundamental problem of strong interaction $P$ and $T$ symmetry and the dark matter problem).   What are the prospects? 

Cosmological measurements could be informative \cite{hw}.   Detection of primordial gravity waves, at attainable sensitivities, would seriously undermine the scenario.  On the other hand, detection of an isocurvature component in the microwave background anisotropy would be very encouraging.  

Recently Arvanitaki and Dubovsky \cite{AD}, elaborating earlier work by themselves and others, have argued that axions whose Compton wavelength is a small multiple of the horizon size of a spinning black hole will form an atmosphere around that hole, populated by super-radiance.  That atmosphere can affect the gravitational wave and x-ray signals emitted from such holes, possibly in spectacular ways.  Since 
\begin{eqnarray}
(m_a)^{-1} ~&\approx&~ 2 \, {\rm cm.} \ \frac{F}{10^{12} \, {\rm GeV}} \nonumber \\
R_{\rm Schwarzschild} ~&\approx&~ 2 \, {\rm km}. \, \ \ \frac{M}{\, M_{\rm Sun}}
\end{eqnarray}
this provides a most promising window through which to view $F \geq 10^{15}$ GeV axions.

Very recently Budker {\it et al}. \cite{bglrs} have proposed ingenious techniques based on nuclear magnetic resonance, that promise to give direct access to a large-$F$ axion background, if such a background indeed supplies the dark matter.   

%%%%%%%%%%%%%%

{\it Acknowledgement}: 
%%%%%%
This work is supported by the U.S. Department of Energy under contract No. DE-FG02-05ER41360.

\end{document}